\documentclass[aps,pra,twocolumn,reprint,showpacs,floatfix,superscriptaddress,longbibliography]{revtex4-1}
\pdfoutput=1
\usepackage[utf8]{inputenc} 
\usepackage{graphicx}
\usepackage{amsmath}
\usepackage{amsfonts}
\usepackage{mathtools}
\usepackage{float}
\usepackage{tabularx, booktabs}
\newcolumntype{Y}{>{\centering\arraybackslash}X}
\newcommand{\be}{\begin{equation}}
\newcommand{\ee}{\end{equation}}
\newcommand{\bea}{\begin{eqnarray}}
\newcommand{\eea}{\end{eqnarray}}

\usepackage{color}

\begin{document}
\title{Two- and three-body problem with Floquet-driven zero-range interactions}
\author{A. G. Sykes}
\author{H. Landa}
\author{D. S. Petrov}
\affiliation{LPTMS, CNRS, Univ.~Paris-Sud, Universit\'{e} Paris-Saclay, F-91405 Orsay, France}

\date{\today}

\begin{abstract}

We study the two-body scattering problem in the zero-range approximation with a sinusoidally driven scattering length and calculate the relation between the mean value and amplitude of the drive for which the effective scattering amplitude is resonantly enhanced. In this manner we arrive at a family of curves along which the effective scattering length diverges but the nature of the corresponding Floquet-induced resonance changes from narrow to wide. Remarkably, on these curves the driving does not induce heating. In order to study the effect of these resonances on the three-body problem we consider one light and two heavy particles with driven heavy-light interaction in the Born-Oppenheimer approximation and find that the Floquet driving can be used to tune the three-body and inelasticity parameters.

\end{abstract}

 \maketitle

\section{Introduction}

Ultra-cold gases provide us with a long list of experimental parameters, tunable {\it in-situ}, providing an unprecedented opportunity to provoke and observe unique dynamical phenomena. One can roughly separate two main directions of research, one of which is associated with the modification of the single-particle dispersion and the other -- with the tuning of the interparticle interactions. The former is a huge field including the development of various trapping techniques, for example via optical potentials~\cite{BlochManyBodyReview,BlochOpticalLattices,BoshierPaintedPotentials} and their time-dependent manipulation, allowing for so-called ``Floquet engineering''.  In particular, significant attention has been on driven many-body lattice Hamiltonians, their dynamics (thermalization or lack there of)~\cite{DAlessioRigolPRX2014,GopalakrishnanKnapDemlerPRB2016,BukovHeylHusePolkovnikovPRB2016,KozarzewskiPrelovsekMierzejewskiPRB2016,RehnLazaridesPollmanMoessnerPRB2016}, and emergent topological phases by inducing artificial gauge fields~\cite{EckardtPRL2005,LignierPRL2007,KitagawaPRB2010,StruckPRL2012,BlochDalibardNascimbene2012,ParkerNatPhys2013,JotzuNature2014,ReichlMuellerPRA2014,GoldmanDalibardPRX2014,RudnerNathanNJP2015,BilitewskiCooperPRA2015,GoldmanDalibardCooperPRA2015,FlaschnerScience2016,NagerlDaleyPRL2016,PlekhanovRouxLehurPRB2017}.



The other direction of research is based on the tunability of interactions via a Fano-Feshbach resonance~\cite{ChinFeshbachResonance} where the scattering length can be changed by several orders of magnitude by varying the strength of a magnetic field. Significant attention has been devoted to dynamical phenomena associated with rapid tuning of the interaction strength. In particular, the (magneto-) association of Feshbach molecules via sweeping the magnetic field has been considered~\cite{Donley2002,RegalNature2003,BorcaBlumeGreene2003,GoralGardiner2004,KohlerJulienneReview2006}. An early experimental paper~\cite{ThompsonWieman2005} exposed ultracold $^{85}$Rb to an oscillating magnetic field for up to 38 ms, with modulation frequencies ranging from 2--9 kHz, and amplitude ranges from 130--280 mG. A clear resonant frequency was observed near the energy of the Feshbach molecule at which atom loss was maximised. Since then, the method involving a modulated magnetic field was applied in a number of different scenarios to produce both homo- and heteronuclear molecules and measure scattering lengths through molecular binding energies~\cite{PappWieman2006,GaeblerStewartBohnJin2007,ZirbelJin2008,WeberInguscioMinardi2008,LangeGrimmChin2009,GrossKokkelmansKhaykovich2011,DykePollackHulet2013}. Modulation frequencies were reported as high as 1 MHz~\cite{DykePollackHulet2013}. Recently a theory was published which explains magneto-association in the case of an oscillating magnetic field~\cite{LangmackHudsonSmithBraaten2015}. This extended a previous work~\cite{HannaKohlerBurnett2007} to include many-body effects in a perturbative treatment involving Tan's contact~\cite{Tan12008}. Finally, a recent paper~\cite{HudsonSmith} has demonstrated how the elastic two-body scattering can be enhanced by an oscillating field at resonance with a weakly-bound molecular state, an issue which was apparently not previously understood. For weak field modulations this method is conceptually similar to the usual magnetic \cite{ChinFeshbachResonance}, optical \cite{FedichevOpticalFeshbach1996,BohnJulienne1997,BohnJuliennePRA1999,PLettPRL2000,TheissGrimmDenschlagPRL2004,ThalhammerGrimmDenschlagPRA2005}, or microwave \cite{PapoularShlyapnikovDalibard} Feshbach-resonance phenomenon which implies coupling of an open-channel continuum to a single quasi-stationary state in another (closed) channel. The difference emerges when the amplitude of the modulation becomes large and one can no longer clearly distinguish the continuum from the molecular state.  

In this paper we solve the two-body scattering problem in the zero-range approximation with periodically driven scattering length by making use of well-established methods~\cite{Shirley1965,SambeQuasienergy,LiReichl1999} that reduce the problem down to a set of recurrence relations involving scattering amplitudes of Floquet channels. Specifically for a sinusoidal drive $a(t)=a_0+2 a_1\cos(\Omega t)$ we find that the steady-state scattering solution is possible for $|a_1|<|a_0|/2$ and the effective scattering length is resonantly enhanced along a family of curves where $a_1$ is a certain function of $a_0$. We find that the Floquet-induced resonances corresponding to different points on these curves are characterized by different effective range parameters. In particular, when $a_0>0$ and in the perturbative limit, $a_1\ll a_0$, these resonances occur when the bound state energy of the Feshbach dimer, $\hbar^2/ma_0^2$, is a positive integer multiple of the driving energy $\hbar\Omega$~\cite{HudsonSmith}. We find, however, that this Floquet-induced resonance is very narrow; the corresponding effective range parameter diverges as $1/a_1^2$. By contrast, this parameter decreases as $1/a_1$ in the opposite limit where both $a_0$ and $a_1$ are large compared to the drive length scale $L_\Omega=\sqrt{\hbar/2\mu\Omega}$ (here $\mu$ is the reduced mass). The Floquet mechanism can thus be used as a tool to tune the resonance width. Remarkably, on these resonances we find no heating due to the absorption of drive quanta.

Another question that we discuss is how the Floquet-induced resonant enhancement of the two-body interaction influences traditional three-body loss processes, catastrophically enhanced near a usual Feshbach resonance~\cite{PappPinoWildJinCornell2008,NavonKrauthSalomon2011,RemSalomon2013,Fletcher2013,MakotynCornellJin2014,SykesHazzardBohn2014,EismannChin2016}. 
To this end we consider the three-body system of two identical heavy bosons or fermions and a light atom for which one can use the Born-Oppenheimer approximation~\cite{Efimov,Fonseca1979}. Specifically, we stay on the Floquet-resonant curve (that is, where $a_1$ is a known function of $a_0$) for the heavy-light interaction and compute the effective Floquet-driven three-body and inelasticity parameter. We find that due to a peculiar feature in the Born-Oppenheimer heavy-heavy potential at distances of the order of $L_\Omega$ these quantities can be dramatically modified compared to their undriven values providing a suppression of inelastic loss rates for strongly-interacting mixtures in the Efimovian regime. The mass-ratio scaling indicates that this suppression is stronger in the equal-mass case.

The  rest of the paper is organised as follows. In Sec.~\ref{TwoBodySection} we treat the two-body problem. The zero-range formalism for a general periodic drive is introduced in Subsec.~\ref{TwoBodyFormalism} and in Subsec.~\ref{SingleFrequencyDrive} we consider the case of a single-frequency sinusoidal drive, present our solution for the resonance position, discuss the near-resonance effective-range expansion and inelastic rate. Section~\ref{ThreeBodySection} is devoted to the three-body problem in the heavy-heavy-light system. In Subsec.~\ref{LightAtom} we solve for the light atom energy and in Subsec.~\ref{HeavyAtom} we use it as the potential energy surface for the heavy atoms calculating the drive-induced shifts of the three-body and inelasticity parameters. In Sec.~\ref{ConclusionsSection} we present our conclusions and discuss avenues of further research.

\section{The two body problem}
\label{TwoBodySection}

\subsection{General Formalism}\label{TwoBodyFormalism}

We start with the Schr\"odinger equation for the relative motion of two particles with reduced mass $\mu$ interacting by a time-dependent interaction potential $V(\mathbf{r},t)$
\begin{equation}\label{schrodinger1}
 i\hbar\partial_t\psi(\mathbf{r},t)=-\frac{\hbar^2}{2\mu}\nabla^2\psi(\mathbf{r},t)+V(\mathbf{r},t)\psi(\mathbf{r},t).
\end{equation}

We are interested in the solution to this problem under the following conditions:
\begin{enumerate}
 \item We assume that the potential can be substituted by the time-dependent Bethe-Peierls boundary condition
 \begin{equation}\label{bethepeierls}
  \lim_{r\rightarrow 0}\frac{\partial_r[r\psi]}{r\psi}=-\frac{1}{a(t)},
 \end{equation}
 which is valid if the relevant length scales of the problem, the de Broglie wave lengths and drive length $L_\Omega$, are much larger than the range of the potential. In atomic physics the latter is given by the van der Waals length $r_{\rm vdW}$. The condition $L_\Omega\gg r_{\rm vdW}$ ensures that the drive is adiabatic on the scale of the potential and its only effect is to change the logarithmic derivative of the wave function just outside the potential range. In other words, $a(t)$ is the instantaneous scattering length corresponding to $V(\mathbf{r},t)$ at a given time $t$. In practice, $a(t)$ is controlled within an experiment through the magnetic field.
 
 \item The time dependence of the potential is periodic, with a fundamental angular frequency $\Omega$, which can be stated equivalently as
 \begin{equation}\label{a}
  a(t)=\sum_{n\in \mathbb{Z}}a_n e^{-in\Omega t}
 \end{equation}
 where $\mathbb{Z}$ denotes the integers and $a_n$ are known constants which we will call the drive parameters. The condition ${\rm Im}a(t)=0$ implies $a_n=a_{-n}^*$.
 
 \item We search for steady-state solutions of these equations. Questions related to quench-dynamics and far-from-equilibrium behavior lie beyond the scope of our current work.
 
 \end{enumerate}
 
In what follows we measure time in units of $1/\Omega$, energy -- in units of $\hbar\Omega$, and distances -- in units of $L_\Omega=\sqrt{\hbar/2\mu\Omega}$. 
 
The scattering solution of Eq.~(\ref{schrodinger1}) can be found by considering the following ansatz
\begin{equation}\label{ansatz1}
 \psi(\mathbf{r},t)=\psi_{\rm in}+\sum_{n\in\mathbb{Z}}f_n\frac{\exp(ik_nr-i\omega_nt)}{r},
\end{equation}
where the incoming wave $\psi_{\rm in}=\exp\left[i\mathbf{k}\cdot\mathbf{r}-i\omega t\right]$ is a solution to the Schr\"odinger equation without interactions, which implies that $\omega=k^2$. The summation in Eq.~\eqref{ansatz1} runs over all the different outgoing scattering channels, labelled by $n$ (occasionally referred to as Floquet sidebands~\cite{LiReichl1999,BagwellLakePRB1992}). Each individual channel satisfies $\omega_{n}=\omega+n=k_{n}^2$ and is preceded by its own individual scattering amplitude $f_n$. We use the convention that $k_n=\sqrt{\omega+n}$ is positive for $\omega+n>0$ corresponding to outgoing waves, whereas channels with $\omega+n<0$ are closed and $ik_n=-\sqrt{-\omega-n}$ is real and negative. The quantity $\omega$ plays the role of {\it quasi-energy} ~\cite{SambeQuasienergy,ZeldovichJETP1967,BilitewskiCooperPRA2015} (occasionally referred to as the Floquet characteristic exponent). Usually, this number can be defined modulo 1 reflecting uncertainty in the number of drive quanta in the system. However, in our case we firmly associate $\omega$ with the well-defined incoming momentum ${\bf k}$ which, in particular, allows us to distinguish two values of $\omega$ whose difference is one. Note further that the interaction potential, when approximated by the boundary condition in Eq.~\eqref{bethepeierls}, only couples the incoming wave with outgoing $s$-wave channels. Hence the addition of higher-order partial-wave channels would not lead to anything interesting under our current set of conditions.

To determine the correct values of $f_n$ we insert Eq.~\eqref{ansatz1} into the Bethe-Peierls condition \eqref{bethepeierls} using Eq.~\eqref{a}. This yields
\begin{align}
 \sum_{n\in\mathbb{Z}}\Bigg[e^{-in t}\left( a_n e^{-i\omega t}+\sum_{p\in\mathbb{Z}}i a_{n-p} k_pf_pe^{-i\omega_pt}\right)+  \nonumber \\
 f_ne^{-i\omega_nt}\Bigg]=0.\label{recurrence1}
\end{align}
Gathering together the terms that share a common prefactor of $e^{-i(\omega+n)t}$ we arrive at a set of recurrence relations~\cite{Shirley1965} that completely determine $f_n$ in terms of the energy of the incoming wave and the drive parameters,
\begin{equation}\label{recurrence2}
f_n+\sum_{p\in\mathbb{Z}}ik_pf_pa_{n-p}=-a_n,
\end{equation}
where $n\in\mathbb{Z}$. 
As a straightforward check, we note that these recurrence relations give the correct result when the scattering length is time independent: That is, with drive parameters $a_n=\delta_{0n}a_0$ ($\delta_{kn}$ being the Kronecker delta), we get $f_n=-\delta_{0n}a_0/(1+ia_0 k)$. This corresponds exactly to the expected scattering amplitude for a zero-range potential with scattering length $a_0$.

An important condition that we require for the solution of Eq.~\eqref{recurrence2} is that $f_n$ decay with increasing $|n|$. Otherwise, we would have to face obvious mathematical and physical difficulties. In particular, it may take infinite time and energy to reach the steady-state. 

As we will see, the amplitudes $f_n$ decay exponentially with $|n|$. In this case the scattering problem that we consider can be treated in much the same manner as the usual one in the presence of finite number of inelastic channels \cite{LandauLifshitz1987}. Then, following the standard terminology, the channel $n=0$ is elastic, since the kinetic energy of the scattered atoms is the same as the incoming one, channels with $\omega+n<0$ are closed, and all other channels are inelastic. The latter can be subdivided into exothermic ($n>0$) and endothermic ($n<0$) channels depending on whether drive quanta are absorbed or emitted. The standard scattering theory \cite{LandauLifshitz1987} relates the total cross section $\sigma_{\rm tot}$, elastic cross section $\sigma_{\rm e}$, and inelastic cross section $\sigma_{\rm r}=\sigma_{\rm tot}-\sigma_{\rm e}$ to the scattering amplitude $f_0$ as
\begin{align}
&\sigma_{\rm tot}=(4\pi/k) {\rm Im}f_0,\label{SigmaTot}\\
&\sigma_{\rm e}=4\pi |f_0|^2.\label{SigmaEl}
\end{align}
The unitarity condition for the scattering, which implies the balance of the incoming and outgoing fluxes of particles through any closed surface, leads to the optical theorem
\begin{equation}\label{optical}
{\rm Im} f_0 = \sum_{n>-\omega}|f_n|^2 k_n.
\end{equation}

Finally, introducing the scattering phase shift $\delta$ related to $f_0$ by
\begin{equation}\label{fthroughdelta}
f_0 = \frac{1}{k\cot\delta-ik},
\end{equation}
one can write the inelastic cross section in the form
\begin{equation}\label{sigmathroughdelta}
\sigma_{\rm r} = -4\pi|f_0|^2{\rm Im}\cot\delta,
\end{equation}
which shows that the appearance of an imaginary part of $\delta$ is related to the presence of inelastic scattering. 

It is important to note that the Floquet scattering theory and standard inelastic scattering theory are equivalent only if a momentum (or energy) detection is implied. Otherwise, the fact of observing a scattered atom pair does not give information on the channel index $n$ and one cannot even distinguish elastic from inelastic scattering. Whether the momentum is detectable depends on a particular experimental realization. For example, in a cold gas elastic collisions are responsible for thermalization and inelastic -- for heating and loss (from a shallow trap). On the other hand, there may be interesting effects if the momentum detection is deliberately avoided such that one can probe interferences between channels. In particular, the flux of atoms through a given surface and probability to find them in a given volume are time dependent due to these interferences. In this case the total scattered flux depends on time and on the surface through which it is measured. Therefore, even the total scattering cross section is then an ambiguous quantity. However, by averaging the flux over the drive period one arrives at Eqs.~(\ref{SigmaTot}) and (\ref{optical}) independent of the surface choice.

\subsection{Single Frequency Drive}\label{SingleFrequencyDrive}

We now consider in detail the case of three non-zero real drive parameters such that
\begin{equation}\label{aSingleFrequency}
a(t)=a_0+2a_1\cos t,
\end{equation}
where without loss of generality we can choose $a_1>0$. For the driving (\ref{aSingleFrequency}) the recurrence relations (\ref{recurrence2}) explicitly read
\begin{equation}
ia_1k_{n+1}f_{n+1}+(1+ia_0k_n)f_n+ia_1k_{n-1}f_{n-1}=-a_n\label{rr1}
\end{equation}
where $n\in\mathbb{Z}$, $a_{-1}=a_1$, and $a_n=0$ for $|n|>1$. Equations~(\ref{rr1}), supplemented with the condition that $f_n$ decay at $|n|\rightarrow \infty$, completely characterize the steady-state scattering solution and can be solved numerically for given values of $a_0$, $a_1$, and $\omega$. As explained in Sec.~\ref{TwoBodyFormalism}, the knowledge of $f_0$ suffices for characterizing the most relevant scattering observables. We have performed a numerical analysis of the problem and calculated the total cross section $\sigma_{\rm tot}$ as a function of $a_0$ and $a_1$ for various collision energies $\omega$. Solutions with decaying $f_n$ can be found only for $a_1<|a_0|/2$. For positive $a_0$ we observe a family of curves, $a_1$ as a function of $a_0$, where the cross section is resonantly enhanced. These curves are shown as solid lines in Fig.~\ref{fig2}. We find that in the limit $\omega\rightarrow 0$ the value of $\sigma_{\rm tot}$ at these resonances tends to infinity $\propto 1/\omega$ as for the usual static case with infinite scattering length.

For weak driving ($a_1\ll 1$) these Floquet resonances appear each time the scattering threshold matches the molecular energy $-1/a_0^2$ plus an integer number of drive quanta, i.e., each time $a_0=1/\sqrt{\nu}$, where $\nu$ is an integer. With increasing $a_1$ the resonances shift and become wider. 

In order to understand and better characterize elastic and inelastic scattering properties near these resonances, let us make the transformation 
\begin{equation}\label{phi}
\phi_n=i(-1)^n k_nf_n,
\end{equation}
with the help of which Eq.~(\ref{rr1}) becomes
\begin{equation}\label{LatticeSchr}
(-\hat\Delta+U_n-\epsilon)\phi_n=\delta_{1|n|}-\delta_{0n}a_0/a_1.
\end{equation}
Equation~(\ref{LatticeSchr}) is a stationary one-dimensional lattice Schr\"odinger equation with the kinetic energy operator defined as lattice Laplacian $\hat\Delta\,\phi_n=\phi_{n-1}-2\phi_n+\phi_{n+1}$, on-site potential $U_n=1/ia_1k_n=-1/a_1\sqrt{-n-\omega}$, energy $\epsilon=2-a_0/a_1$, and a source term on the right-hand side. The first thing to mention is that in the limit $|n|\rightarrow \infty$ the potential can be neglected and there are two types of solutions for $\phi_n$: plane waves when $\epsilon$ is inside the band $0<\epsilon<4$ (i.e., $-2<a_0/a_1<2$) or exponentially decaying or growing functions, if $\epsilon$ is outside the band. We are interested in the latter case (particularly, $\phi_n$ should exponentially decay with $|n|$) and thus are limited to the domain $a_1<|a_0|/2$.

We now discuss the limit $\omega\rightarrow 0$. The on-site potential in this case is real and attractive for $n<0$, purely imaginary for $n>0$, and $U_0$ diverges as $\propto 1/\sqrt{\omega}$. If not exactly at resonance (see Subsec.~\ref{Subsec:ResPos}), one can first set $\phi_0=0$, which effectively decouples Eqs.~(\ref{LatticeSchr}) into the positive-$n$ part
\begin{equation}\label{SchrPlus}
[-\hat\Delta_+ + U_n^{(0)} - \epsilon]\phi_n=\delta_{1n},\; \; n>0,
\end{equation}
and the negative-$n$ part
\begin{equation}\label{SchrMinus}
[-\hat\Delta_- + U_n^{(0)} - \epsilon]\phi_n=\delta_{-1n},\;\;  n<0,
\end{equation}
where $U_n^{(0)}=U_n(\omega=0)$ and we have introduced the positive-side operator $\hat\Delta_+$ equal to $\hat\Delta$ for $n>1$ and defined as $\hat\Delta_+ \phi_1 = -2\phi_1+\phi_2$ for $n=1$. Similarly, on the negative side $\hat\Delta_- \phi_n=\Delta \phi_n$ for $n<-1$ and $\hat\Delta_- \phi_{-1} = \phi_{-2}-2\phi_{-1}$.

If Eqs.~(\ref{SchrPlus}) and (\ref{SchrMinus}) are not singular, their solutions provide us with finite values for $\lim_{\omega\rightarrow 0}\phi_1$ and $\lim_{\omega\rightarrow 0}\phi_{-1}$. We can then go back to Eq.~(\ref{LatticeSchr}) for $n=0$ writing it as
\begin{equation}\label{phi0}
\phi_0=\frac{a_1(\phi_{-1}+\phi_1)-a_0}{a_0+1/i\sqrt{\omega}},
\end{equation}
which {\it a posteriori} confirms that $\phi_0$ vanishes at $\omega=0$. By using Eq.~(\ref{phi}) to express $f_0$ through $\phi_0$ and taking the limit $\omega\rightarrow 0$ in Eq.~(\ref{phi0}) we define the effective scattering length
\begin{equation}\label{EffScatLength}
a_{\rm eff}=-\lim_{\omega\rightarrow 0}f_0=a_0-a_1\lim_{\omega\rightarrow 0}(\phi_{-1}+\phi_1).
\end{equation}
Note, that the imaginary contribution to $a_{\rm eff}$ comes only from $\phi_1$, which is complex because of the imaginary potential $U_n$ on the positive-$n$ side. The imaginary part of $a_{\rm eff}$ is thus associated with nonvanishing inelastic scattering amplitudes for $n>0$.

\subsubsection{Resonance positions\label{Subsec:ResPos}}

From Eq.~(\ref{EffScatLength}) one sees that resonances in $a_{\rm eff}$ can only occur when $\phi_1$ or $\phi_{-1}$ diverge. This can happen when the operators on the left-hand sides of Eqs.~(\ref{SchrPlus}) or (\ref{SchrMinus}) are singular (have zero eigenvalues). However, because of the structure of $U_n$ for $n>0$, $\phi_{1}$ is never infinite. In order to see this assume that there is a solution $\phi_n$ satisfying the homogeneous equation $[-\hat\Delta_+ + U_n^{(0)} - \epsilon]\phi_n=0$. Let us multiply this equation by $\phi^*_n$ and sum it over $n$ from 1 to $\infty$. We then perform the summation by parts by using the equality $[-\hat\Delta_+ +U^{(0)*}_n-\epsilon^*]\phi^*_n=0$ and obtain $\sum_{n=1}^{\infty}{\rm Im}[U_n^{(0)}-\epsilon]|\phi_n|^2=0$. Since $\epsilon$ is real, ${\rm Im}U_n^{(0)}=-1/a_1\sqrt{n}$, and $|\phi_n|^2>0$, we conclude that such $\phi_n$ does not exist.

\begin{figure}
 \includegraphics[width=8cm]{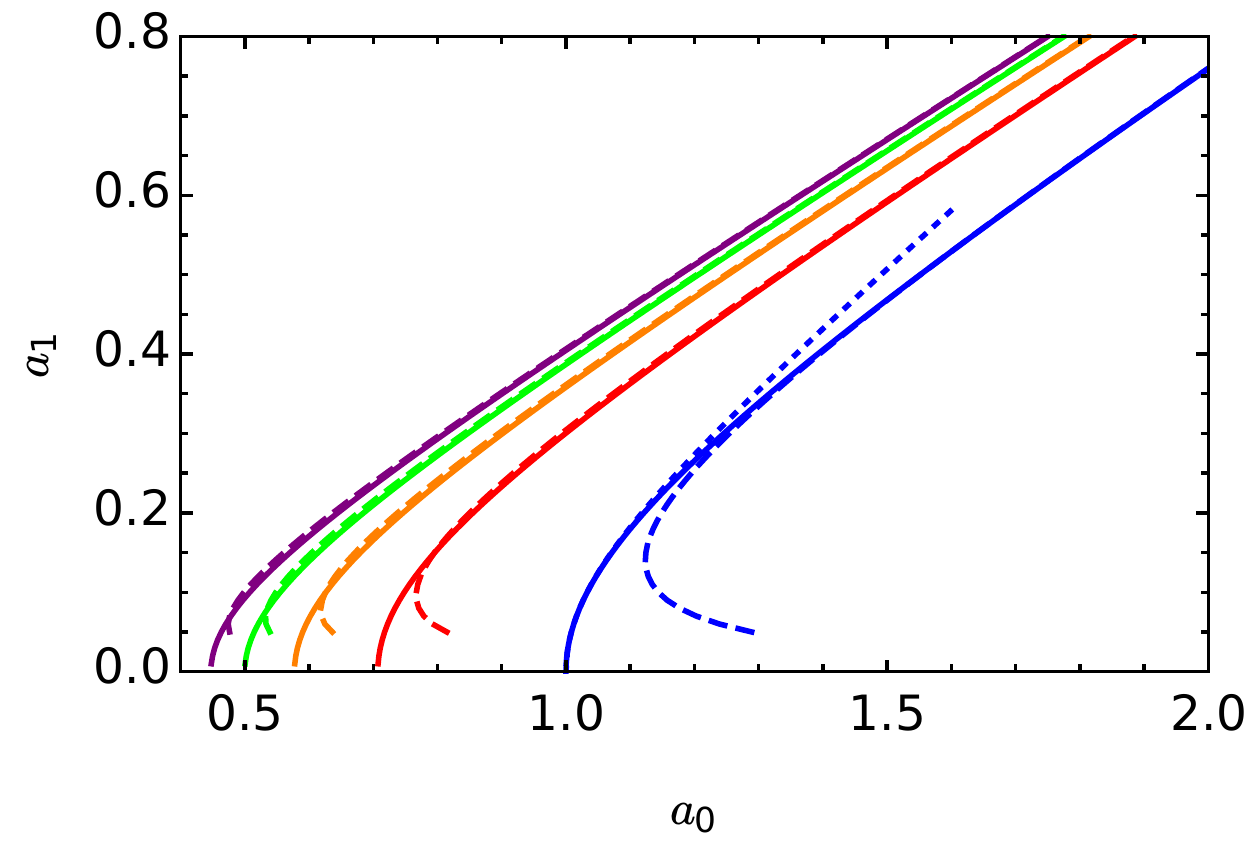}
 \caption{The resonance-position curves for the first five branches in the $\{a_0,a_1\}$ plane (solid). The dashed lines show the large-$a_1$ asymptotes \eqref{analytic_resonances} with $q$ given after Eq.~(\ref{transcendental}). The dotted blue curve shows the result of Eq.~\eqref{HudsonSmithLimit} valid for small $a_1$. The resonance curves intersect with the $a_1=0$ axis at $a_0=1/\sqrt{\nu}$, where the (undriven) bound-state energy equals $\nu=1,2,3,\ldots$ drive quanta.}
 \label{fig2}
\end{figure}

Resonances that we observe in Fig.~\ref{fig2} appear for $\phi_{-1}=\infty$ when $\epsilon$ exactly corresponds to the energy of a bound state in the potential $U_n^{(0)}=-1/a_1\sqrt{-n}$ for negative $n$. The eigenvalue problem reads
\begin{equation}\label{SchrMinusHom}
 [-\hat\Delta_- + U_n^{(0)} - \epsilon_\nu]\phi^{(\nu)}_n=0
\end{equation}
and, because of the fat tail of the inverse-square-root potential, it has an infinite number of bound solutions (which we label by $\nu=1,2,...$) with the accumulation point $\epsilon_{\nu\rightarrow\infty}\rightarrow -0$. We choose $\phi^{(\nu)}_n$ to be square-normalized eigenfunctions and we point out that the $\epsilon_{\nu}$ depend only on $a_1$. The equation $\epsilon=2-a_0/a_1=\epsilon_\nu$ for different $\nu$ gives a family of resonance curves in the $\{a_0,a_1\}$ plane,  which accumulate at $a_1=a_0/2$. In the limit $a_1\rightarrow 0$ the operator $\hat\Delta_-$ in Eq.~(\ref{SchrMinusHom}) can be neglected and we have $\epsilon_\nu\approx -1/a_1 \sqrt{\nu}$ arriving at the resonance positions $a_0=1/\sqrt{\nu}$ discussed earlier. In this limit $\phi_{n}^{(\nu)}=\delta_{-\nu n}$ and for small but finite $a_1$ the function $\phi_n^{(\nu)}$ exponentially decays with $|n+\nu|$ as $\propto a_1^{|n+\nu|}$ and $\epsilon_\nu$ can be systematically expanded in powers of $a_1$. Technically, this can be done by introducing a cut-off at a finite $|n+\nu|$. In particular, for $\nu=1$, by imposing $\phi_{-4}=0$ in Eq.~(\ref{SchrMinusHom}) one obtains
\begin{equation}\label{HudsonSmithLimit}
a_1\approx\sqrt{\frac{(a_0-1)(2a_0-\sqrt{2})(\sqrt{3}-3a_0)}{2(3+\sqrt{3}-6a_0)}}.
\end{equation}
Equation~(\ref{HudsonSmithLimit}) is shown in Fig.~\ref{fig2} as the blue dotted line. It is straight-forward to derive formulae similar to Eq.~(\ref{HudsonSmithLimit}) for the resonance curves with $\nu>1$.

In the limit of large $a_1$ the characteristic length scale of $U_n$ is proportional to $a_1^{2/3}\gg 1$ and the lattice problem (\ref{SchrMinusHom}) becomes equivalent to solving the continuum Schr\"odinger equation 
\begin{equation}\label{sqrtschrodinger}
\left(-\frac{\partial^2}{\partial x^2}-\frac{1}{a_1 \sqrt{-x}}\right)\phi_x^{(\nu)}=\epsilon_\nu \phi_x^{(\nu)},
\end{equation}  
with the boundary conditions $\phi_{x\rightarrow -\infty}^{(\nu)}=\phi_0^{(\nu)}=0$. This problem is analytically solvable~\cite{LamieuxBose, Ishkhanyan} with the solution for the $\nu$-th bound state given by
\begin{equation}\label{phix}
 \phi_x^{(\nu)}\propto e^{\sqrt{|\epsilon_\nu|}x}u'(z),
 \end{equation}
where we have defined an auxiliary function $u(z)=\exp(\sqrt{2q}z)[\alpha {\rm H}_{q}(-z)+{}_1\nobreak\!{\rm F}_1(-q/2;1/2;z^2)]$ of variable $z=|\epsilon_\nu|^{1/4}\sqrt{-2x}-\sqrt{2q}$, ${\rm H}_q$ is the Hermite function, $_1\nobreak\!{\rm F}_1$ -- the Kummer confluent hypergeometric function~\cite{SpecialFunctionsBook}, the constant $\alpha=[2q\, _1\nobreak\!{\rm F}_1(1-q/2;3/2;2q)+{}_1\nobreak\!{\rm F}_1(-q/2;1/2;2q)]/[\sqrt{2q}{\rm H}_{q-1}(\sqrt{2q})-{\rm H}_q(\sqrt{2q})]$, and the parameter $q=|\epsilon_{\nu}|^{-3/2}/4a_1^2$ is determined by choosing the $\nu$-th smallest root of the transcendental equation
\begin{equation}\label{transcendental}
 \sqrt{2q}{\rm H}_{q-1}(-\sqrt{2q})+{\rm H}_q(-\sqrt{2q})=0.
\end{equation}
The smallest five roots are 0.862318, 1.85141, 2.84706, 3.84463, 4.84306 and the series continues to infinity with the distance between neighboring roots approaching 1. Thus, the resonance curves, $a_1$ as a function of $a_0$, at large $a_1$ (and $a_0$) are given by the implicit equation  
\begin{equation}\label{analytic_resonances}
a_0=(4q)^{-2/3}a_1^{-1/3}+2a_1,
\end{equation}
which demonstrates how the resonances accumulate close to the line $a_1=a_0/2$. Equation~(\ref{analytic_resonances}) is used to generate the dashed lines in Fig.~\ref{fig2}.

\subsubsection{Near-resonance effective-range expansion}

We now discuss the behavior of the elastic scattering amplitude $f_0$ for finite deviation from the resonance and for finite $\omega$. We can still write the set of Eqs.~(\ref{LatticeSchr}) as two sets of equations:
\begin{equation}\label{SchrPlusdelta}
[-\hat\Delta_+ + U_n^{(0)} - \epsilon_\nu +\delta U_n-\delta \epsilon]\phi_n=(1+\phi_0)\delta_{1n}
\end{equation}
and
\begin{equation}\label{SchrMinusdelta}
[-\hat\Delta_- + U_n^{(0)} - \epsilon_\nu +\delta U_n-\delta \epsilon]\phi_n=(1+\phi_0)\delta_{-1n},
\end{equation}
where we have introduced $\delta U_n = U_n - U_n^{(0)}$ and $\delta \epsilon=-\delta a_0/a_1$, which corresponds to a deviation $\delta a_0$ of $a_0$ from its $\nu$-th resonant position at fixed $a_1$. In order to find $f_0$ we follow the same logic as for the case $\phi_0=0$. Namely, we can solve Eqs.~(\ref{SchrPlusdelta}) and (\ref{SchrMinusdelta}), thus expressing $\phi_1$ and $\phi_{-1}$ through $\phi_0$, and substitute the results into Eq.~(\ref{phi0}), which is then a closed equation for $\phi_0$ or $f_0$. Note the simple linear dependence of $\phi_{1}$ and $\phi_{-1}$ on $\phi_0$. Namely, by denoting solutions of Eqs.~(\ref{SchrPlusdelta}) and (\ref{SchrMinusdelta}) with $\phi_0=0$, respectively, by $G_{1n}$ and $G_{-1n}$ we have $\phi_{\pm 1}=(1+\phi_0)G_{\pm 1 \pm 1}$. Substituting these expressions into Eq.~(\ref{phi0}) and using (\ref{phi}) we obtain
\begin{equation}\label{ScatAmplGen}
 f_0=\left[\frac{1}{a_1(G_{11}+G_{-1-1})-a_0}-i\sqrt{\omega}\right]^{-1}.
\end{equation}

The scattering problem is now reduced to finding $G_{\pm 1\pm 1}$ and we can discuss the effective-range expansion of $f_0$ close to the $\nu$-th resonance (i.e., we assume small collision energy $\omega$ and detuning $\delta \epsilon$). Note that $\delta U_n=\omega n^{-3/2}/2ia_1+O(\omega^2)$ for $n>0$ and $\delta U_n=-\omega(-n)^{-3/2}/2a_1+O(\omega^2)$ for $n<0$. 

Equation~(\ref{SchrMinusdelta}) can be solved perturbatively by expanding the solution in the basis of eigenfunctions  $\phi_n^{(\nu)}$ of the operator $-\hat\Delta_- + U_n^{(0)}$ [see Eq.~(\ref{SchrMinusHom}) and we remind of the normalization $\sum_{n=1}^{\infty}|\phi_{-n}^{(\nu)}|^2=1$]. The leading-order result for $G_{-1-1}$ reads
\begin{equation}\label{Gm1}
 G_{-1-1}=\frac{|\phi_{-1}^{(\nu)}|^2}{-\delta\epsilon+\sum_{n=1}^{\infty}\delta U_{-n}|\phi_{-n}^{(\nu)}|^2}
\end{equation}
and the next-order term is $O(1)$.

As we have explained, Eq.~(\ref{SchrPlusdelta}) is never singular (for physically relevant parameters). The quantity $G_{11}$ can thus be approximated by a constant independent of $\omega$ and $\delta\epsilon$ and is comparable to the next-to-leading order contribution to $G_{-1-1}$.

\begin{figure}
    \centering
    \includegraphics[width=8.6cm]{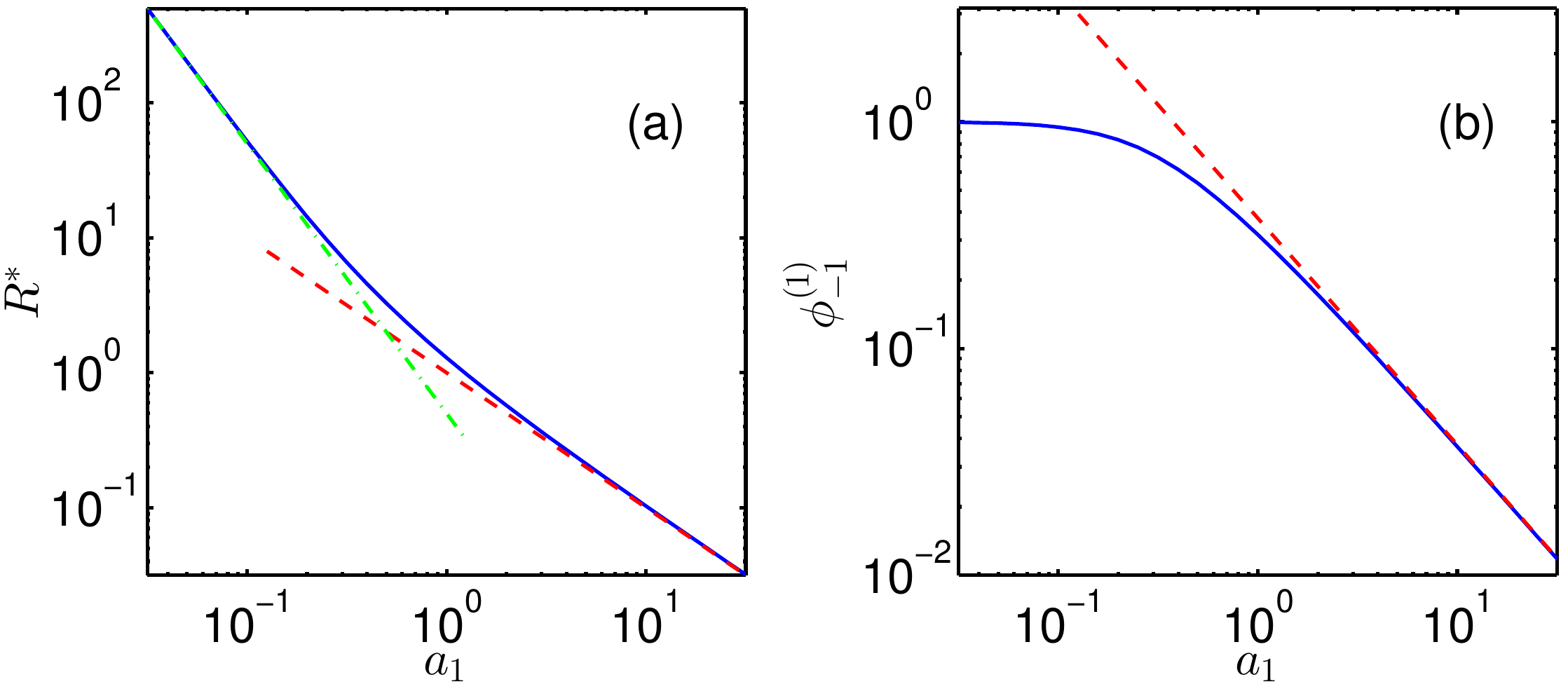}
    \caption{(a) $R^*$ versus $a_1$. The solid curve is calculated numerically using Eq.~\eqref{Rs}. The asymptotic limits $1/2a_1^2$ and $1/a_1$ are shown as the green dash-dotted and red dashed lines, respectively. (b) $\phi_{-1}^{(1)}$ versus $a_1$. This parameter is relevant for the determination of $a_{\rm eff}$ in Eq.~\eqref{aeff}. The red dashed line is given by $\phi_{-1}^{(1)}=0.3754/a_1$. In both panels $a_0$ is a function of $a_1$ such that we stay on the first Floquet resonance curve.}
    \label{fig:Rs}
\end{figure}

Substituting Eq.~(\ref{Gm1}) into Eq.~(\ref{ScatAmplGen}) and keeping only the leading-order terms we obtain
\begin{equation}\label{ScatAmplLeading}
 f_0=-\frac{1}{1/a_{\rm eff}+R^{{}*{}} \omega+i\sqrt{\omega}},
\end{equation}
where the effective scattering length is
\begin{equation}\label{aeff}
 a_{\rm eff}=a_1|\phi_{-1}^{(\nu)}|^2/\delta\epsilon=-a_1^2|\phi_{-1}^{(\nu)}|^2/\delta a_0
\end{equation}
and the width parameter
\begin{equation}\label{Rs}
 R^*=\frac{\sum_{n=1}^{\infty}n^{-3/2}|\phi_{-n}^{(\nu)}|^2}{2a_1^2|\phi_{-1}^{(\nu)}|^2}. 
\end{equation}
The quantities $R^*$ and $\phi_{-1}^{(1)}$ are shown in Fig.~\ref{fig:Rs} as a function of $a_1$ (with $a_0$ varying appropriately to stay on the first resonance curve). 

For small $a_1$ near the first resonance ($\nu=1$) the effective scattering length $a_{\rm eff}\approx -a_1^2/\delta a_0$ and $R^*\approx 1/2a_1^2$ indicating a very narrow resonance. Here we use the fact that  $|\phi_{n}^{(1)}|^2\approx \delta_{-1n}$. In fact, resonances with $\nu>1$ are even narrower since $|\phi_{-1}^{(\nu)}|\propto a_1^{\nu-1}$.  

For large $a_1$ the extent of $\phi_n^{(\nu)}$ as a function of $n$ is approximately $a_1^{2/3}$ (the characteristic length scale of the potential $U_n$) and taking into account the normalization we have $\phi_{-1}^{(\nu)}\sim 1/a_1$. Equations~(\ref{aeff}) and (\ref{Rs}) in this case give $a_{\rm eff}\sim -1/\delta a_0$ and $R^*\sim 1/a_1$ and these scalings stay qualitatively the same when we change $\nu$. A more precise determination of $a_{\rm eff}$ requires to know the normalization constant of the continuum-limit wave function (\ref{phix}). By contrast, $R^*$ can be found more precisely by using the following arguments. The first derivative of Eq.~(\ref{sqrtschrodinger}) with respect to $x$ reads 
\begin{equation}\label{sqrtschrodingerDer}
\left(-\frac{\partial^2}{\partial x^2}-\frac{1}{a_1\sqrt{-x}}-\epsilon_\nu\right)\frac{\partial \phi_x^{(\nu)}}{\partial x}-\frac{\phi_x^{(\nu)}}{2a_1(-x)^{3/2}}=0.
\end{equation}  
We multiply it by $\phi_x^{(\nu)*}$ and integrate over $x$ from $-\infty$ to 0 by using the integration by parts and Eq.~(\ref{sqrtschrodinger}). The result is
\begin{equation}\label{Result}
 \int_0^\infty x^{-3/2}|\phi_{-x}^{(\nu)}|^2 dx = 2a_1|\partial \phi_x^{(\nu)}/\partial x|_{x=0}|^2.
\end{equation}
Note that for large $a_1$ the derivative $\partial \phi_x^{(\nu)}/\partial x$ at $x=0$ equals $-\phi_{-1}^{(\nu)}$. By comparing Eqs.~(\ref{Result}) and (\ref{Rs}) we finally obtain in the large-$a_1$ limit that $R^*=1/a_1$ independent of $\nu$, i.e., the Floquet resonances become wider with increasing $a_1$ and we can use them to control $R^*$.

\subsubsection{Inelastic rate}

Since $a_{\rm eff}$ and $R^*$ are real, Eq.~(\ref{ScatAmplLeading}) tells us that the scattering is elastic in the limit of small $\omega$ and $\delta a_0$ [compare Eq.~(\ref{ScatAmplLeading}) with Eqs.~(\ref{fthroughdelta}) and (\ref{sigmathroughdelta})]. Moreover, $G_{-1-1}$ is real to all orders and the leading contribution to the inelastic scattering cross section is related to the imaginary part of $G_{11}$ as
\begin{equation}\label{sigmar}
\sigma_{\rm r}=\frac{4\pi|f_0|^2}{k}(a_{\rm eff}^{-1}+R^*\omega)^2 a_1 {\rm Im}G_{11}.
\end{equation}
Equation~(\ref{sigmar}) can be discussed for various values of the parameters $\omega$, $a_{\rm eff}$, $R^*$, and $a_1$. For simplicity, let us look at resonance, $a_{\rm eff}=\infty$ and assume that $a_1\sim 1$. Then, $R^*$ and ${\rm Im} G_{11}$ are also of the order of 1. Neglecting $\omega^2$ compared to $\omega$ we obtain $\sigma_{\rm r}\sim \sqrt{\omega}$ and the two-body loss rate constant $\sigma_{\rm r}k\sim \omega$ or, in physical units, $\sigma_{\rm r}v\sim (\hbar L_\Omega/\mu)(kL_\Omega)^2$, where $v$ is the relative velocity. We see that the two-body loss rate in such a thermal Floquet-unitary gas scales with the first power of temperature $T$ compared to the three-body loss rate, which is expected to be proportional to $1/T^2$~\cite{RemSalomon2013,Fletcher2013}.  

\begin{figure}
    \centering
    \includegraphics[width=6cm]{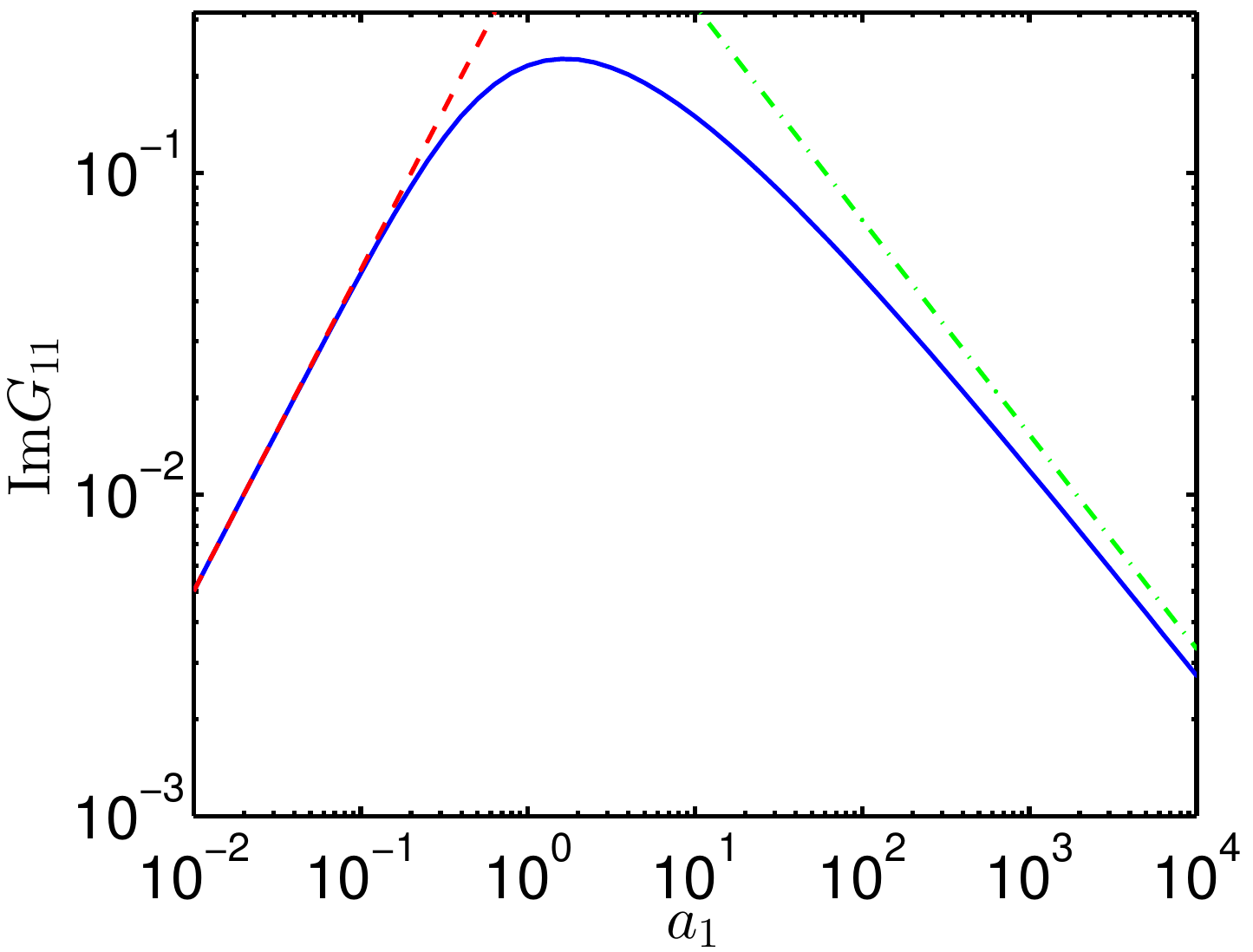}
    \caption{The quantity ${\rm Im}G_{11}$, necessary for characterizing the inelastic cross section (\ref{sigmar}), versus $a_1$ on the first Floquet resonance curve. The small- and large-$a_1$ asymptotes are shown as the red dashed and green dash-dotted lines, respectively.}
    \label{fig:ImG11}
\end{figure}

The quantity ${\rm Im}G_{11}$ is plotted in Fig.~\ref{fig:ImG11} versus $a_1$ (with $a_0$ staying on the first resonance curve) as a solid curve. We also show its small (red dashed) and large (green dash-dotted) asymptotes. The former is calculated from Eq.~(\ref{SchrPlusdelta}) by neglecting the operator $\hat\Delta_+$. Namely, ${\rm Im}G_{11}\approx \nu a_1/(1+\nu)$, where we have used $a_0\approx 1/\sqrt{\nu}$. In the opposite limit $a_1\gg 1$ the potential $U_n$ and $\epsilon_\nu\approx -(4q)^{-2/3}a_1^{-4/3}$ [recall $q$ is the $\nu$-th smallest root of Eq.~\eqref{transcendental}] comprise the perturbation. Then, in the first approximation $G_{11}$ is real and in the next order we obtain ${\rm Im}G_{11}=\sqrt{\pi/2}a_1^{-1}(-\epsilon_{\nu})^{-1/4}\propto a_1^{-2
/3}$. 

\section{The Three-Body Problem}\label{ThreeBodySection}

Three particles with resonant pair-wise interactions can exhibit peculiar effects which go under the name of Efimov physics \cite{Efimov}. 
In particular, the complete description of an Efimovian three-body system requires a three-body parameter (or three-body phase) which absorbs the short-range three-body physics in the same manner as the scattering length absorbs the short-range two-body physics. The three-body phase fixes positions of Efimov trimer states and relates them to all other three-body observables. 

In the case of atoms, when three of them approach each other to distances of the order of $r_{\rm vdW}$, they can recombine forming a deeply-bound molecular state. The molecular binding energy released in the form of the molecule-atom recoil kinetic energy is much higher than the temperature and the trap depth and the products of the reaction are lost. This is one of the main mechanisms strongly shortening life times of resonant Bose gases and mixtures. In the zero-range approximation this type of local three-body loss processes can be modeled by adding an imaginary part to the three-body phase (also called inelasticity parameter). This is in analogy with the two-body losses modeled by an imaginary part of the two-body scattering phase shift. The three-body loss is resonantly enhanced when an Efimov trimer passes through the three-atom threshold. The real part of the three-body phase controls the position of this resonance and its imaginary part -- the overall scale and width. 

While the two-body scattering length can be modified by the magnetic field, the three-body phase is not so easily tunable. In this section we analyse the possibility to tune it by going along the Floquet resonance curve where $a_{\rm eff}=\infty$. We have already mentioned that along this curve the width of the two-body resonance changes. The effect of the resonance width and the appearance of the characteristic length scale $R^*$ in the Efimov three-body problem has been discussed in Refs.~\cite{Petrov2004,GogolinMoraEgger2008,WangDIncaoEsry2011,SchmidtRathZwerger2012}. Here we are interested in the behavior of the three-body system at distances of the order of $L_\Omega$ (one in our units). In order to get an insight into this problem we consider the system of a light atom and two identical heavy atoms (fermions or bosons), where the heavy-light interaction is Floquet resonant. This set-up allows us to work in the Born-Oppenheimer approximation considering first the light-atom problem in the field of fixed heavy scatterers and then using the light-atom energy as the potential energy surface for the heavy atoms.

\subsection{The light-atom problem}\label{LightAtom}

\begin{figure}[t!]
\includegraphics[width = 8cm]{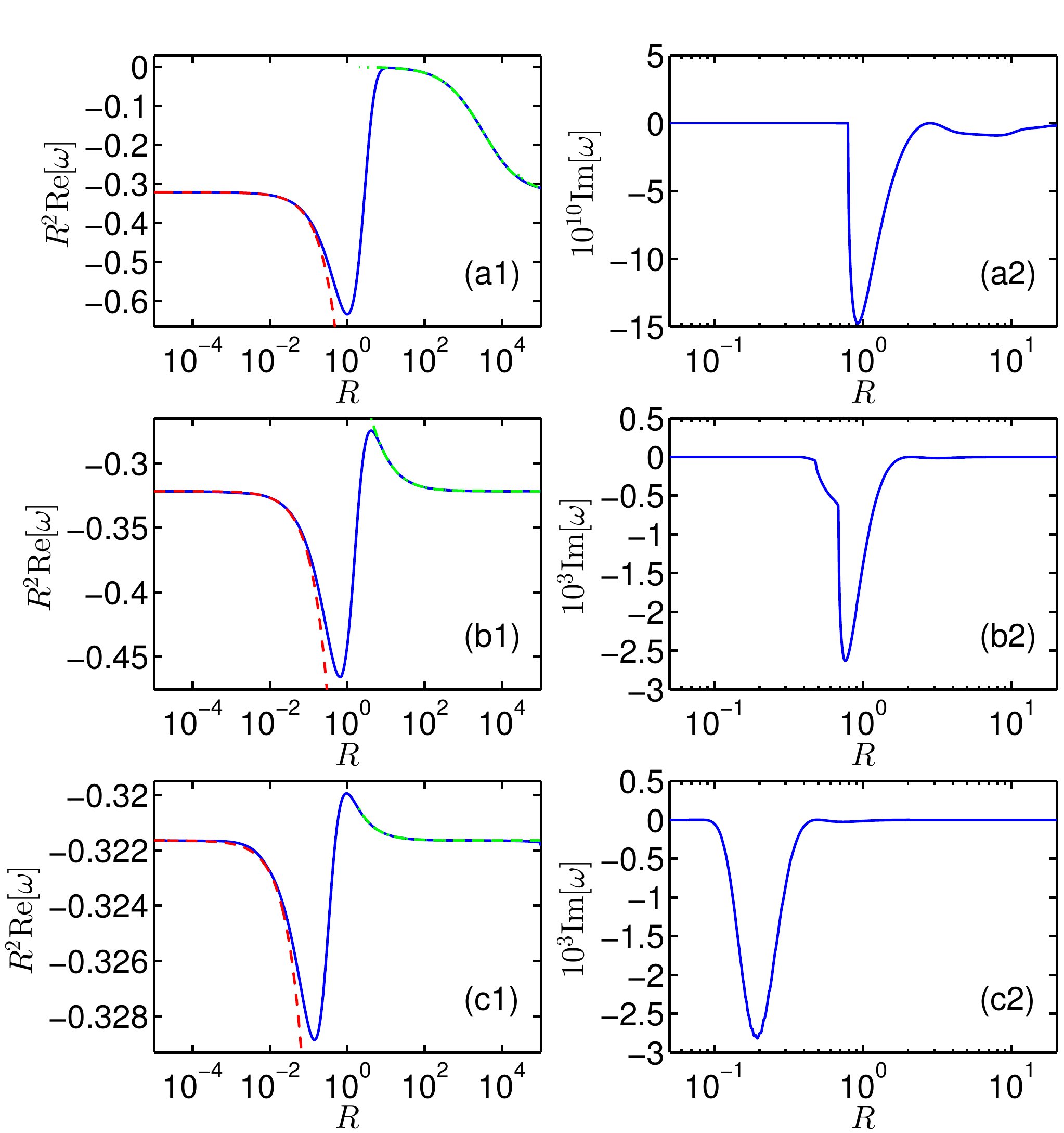}
 \caption{The energy of the light particle as a function of the distance between the heavy particles for the drive parameters on the first resonance curve (rightmost in Fig.~\ref{fig2}). Subfigures (a), (b), and (c) stand, respectively, for $a_1=0.01,1$, and $100$. We observe a single undulation of Re($\omega$)$R^2$ when the heavy particles are separated by the length scale of the drive. The green dash-dotted curves correspond to solutions of Eq.~(\ref{BOundrivenRstar}) and the red dashed curves are the short-$R$ asymptotes (\ref{omegasmallR}). The right column shows the imaginary part of $\omega(R)$, which is always significantly smaller than the real part.
 }
 \label{fig4}
\end{figure}

Let us consider an atom of mass $m=\mu$ interacting with two infinitely-heavy scatterers and model this system by the Schr\"odinger equation (\ref{schrodinger1}) where $V({\bf r},t)$ is substituted by $V({\bf r}-{\bf R}/2,t)+V({\bf r}+{\bf R}/2,t)$. Here we are in a reference frame where the heavy particles are located at ${\bf R}/2$ and $-{\bf R}/2$, and therefore separated by $R$. We make the same assumptions concerning the zero-range approximation and steady state solution as in Sec.~\ref{TwoBodySection}. In addition, we assume the single frequency drive and that we are exactly on the first resonant curve. However, this time we will not look for the scattering solution with finite incoming wave $\psi_{\rm in}$ at a given $\omega>0$, but rather a bound state at negative ${\rm Re}\omega$ which does not contain the incoming wave. Accordingly, we adopt the ansatz
\begin{equation}\label{ansatz2}
\psi(\mathbf{r},t)=\sum_{n\in\mathbb{Z}}f_n\sum_{\pm}\frac{\exp\left[ik_n|{\bf r}\pm{\bf R}/2| -i\omega_nt\right]}{|{\bf r}\pm{\bf R}/2|},
\end{equation}
where we use the same notation $\omega_n=\omega +n = k_n^2$ and the same convention for choosing the branch of square root for calculating $k_n$. By applying the Bethe-Peierls boundary condition (\ref{bethepeierls}) either at ${\bf r}\rightarrow {\bf R}/2$ or at ${\bf r}\rightarrow -{\bf R}/2$ and by grouping terms with the same time dependence, we arrive at the recurrence relations [cf. Eq.(\ref{rr1})]
\begin{equation}
ia_1\tilde{k}_{n+1}f_{n+1}+(1+ia_0\tilde{k}_n)f_n+ia_1\tilde{k}_{n-1}f_{n-1}=0,\label{rr2}
\end{equation}
where $i\tilde{k}_n=ik_n+\exp(ik_nR)/R$. We are interested in calculating $\omega(R)$ for which Eq.~(\ref{rr2}) has a solution under the requirement that $f_n$ decay at large $|n|$. In addition, in the limit $R\rightarrow \infty$ the solution should correspond to the two-body solution at zero collision energy obtained in Sec.~\ref{TwoBodySection}. Our numerical results are shown in Fig.~\ref{fig4} where we plot ${\rm Re}\omega(R)R^2$ and ${\rm Im}\omega(R)$. Visible kinks in the curves ${\rm Im}\omega(R)$ happen each time ${\rm Re}\omega(R)$ is close to an integer corresponding to the opening of a new inelastic channel. Note, however, that the imaginary part of $\omega(R)$ is always much smaller than its real part, which, as one can see, has a distinct feature at $R\sim 1$. 

Let us now discuss physical origins of this behavior. It is instructive to first mention what happens with the undriven problem at a constant scattering length $a$. This case is described by setting $a_0=a$ and $a_1=0$ in Eq.~(\ref{rr2}). The solution is $f_n=\delta_{0n}$ and $\omega$ follows from the equation $1+ia\tilde{k}_0=0$ which explicitly reads
\begin{equation}\label{BOundriven}
\sqrt{-\omega}R=\exp(-\sqrt{-\omega}R)+R/a.
\end{equation}
When $R\ll |a|$ (or for all $R$ when $a=\infty$) we can neglect the last term in the right hand side of Eq.~(\ref{BOundriven}) and arrive at the well-known inverse-square scaling $\omega(R)=-C^2/R^2$, where $C\approx 0.567$ is the solution of $\exp(-C)=C$. We expect this scaling to occur in the driven problem when we are exactly on the Floquet resonance ($a_{\rm eff}=\infty$), at least,  for sufficiently large $R$ where we can neglect the interference effects of the closed-channel evanescent waves (with $n<0$) and consider the light-heavy interaction as if it is undriven and characterized by an infinite scattering length. Indeed, this is what we observe in Fig.~\ref{fig4}. 

This logic can be continued and we can explain the decaying part of the curve ${\rm Re}\omega(R)R^2$ at large $R$ by including the effective-range term in Eq.~(\ref{ScatAmplLeading}) into account. Note that the two-body scattering amplitude (\ref{ScatAmplLeading}) can be derived for the undriven case by inserting the energy-dependent scattering length $1/a(\omega)=1/a_{\rm eff}+R^*\omega$ in the Bethe-Peierls boundary condition. Equation~(\ref{BOundriven}) for $a_{\rm eff}=\infty$ then becomes
\begin{equation}\label{BOundrivenRstar}
\sqrt{-\omega}R=\exp(-\sqrt{-\omega}R)+RR^*\omega 
\end{equation} 
the solution of which is shown in Fig.~\ref{fig4} as the dashed green curves. We clearly see that $\omega(R)$ for $R\gg 1$ can well be explained by solving the undriven problem (\ref{BOundrivenRstar}) with light-atom interaction characterized by the scattering amplitude (\ref{ScatAmplLeading}). If, in addition to $R\gg 1$, we have $R\gg R^*$ (these conditions are not always equivalent), Eq.~(\ref{BOundrivenRstar}) can be solved perturbatively with the result 
\begin{equation}\label{omegalargeR}
\omega\approx -\frac{C^2}{R^2}+\frac{2C^3}{1+C}\frac{R^*}{R^3}.
\end{equation}

Equation~(\ref{BOundriven}) can also be used to explain the small-$R$ behavior in the driven case observed in Fig.~\ref{fig4}. Indeed, in this limit the energy of the light atom is much larger than 1, which means that it adjusts itself adiabatically to $a(t)$ and its instantaneous energy can be found from the equation $\sqrt{-E(t)}R=\exp[-\sqrt{-E(t)}R]+R/a(t)$. The Floquet quasi-energy $\omega$ is then (by definition) the average of $E(t)$ over the drive period. Expanding at small $R/a(t)$ and averaging we obtain
\begin{equation}\label{omegasmallR}
\omega\approx -\frac{C^2}{R^2}-\frac{2C}{1+C}\frac{1}{\sqrt{a_0^2-4a_1^2}R}.
\end{equation}
In Fig.~\ref{fig4} this scaling is shown as the red dashed curves. The second term in the right hand side of Eq.~(\ref{omegasmallR}) gives a negative correction to the simple $\omega\sim 1/R^2$ scaling. Physically, this reflects a stronger attraction of the light atom to the heavy ones since, compared to the unitary case ($a=\infty$), the heavy-light potentials characterized by a finite positive scattering length are more attractive. We should mention that one has to be careful in the case of large $a_1$ where, over a short time period, $a(t)$ is close to its smallest value $a_0-2a_1\propto 1/a_1^{1/3}\ll 1$ [see Eq.~(\ref{analytic_resonances})]. Therefore, our derivation of Eq.~(\ref{omegasmallR}) is valid for $R\ll a_1^{-1/3}$ which is more strict than $R\ll 1$.

To summarize this section we find that exactly on the Floquet resonance  the small- and large-$R$ asymptotes of the light-atom energy scale as $\omega(R) \propto 1/R^2$. The motion of the heavy atoms in this inverse-$R^2$ potential can lead to the phenomenon called the fall of a particle to the center or, equivalently, to the Efimov effect. The new physics occurs when the heavy particles are separated by approximately the length-scale of the drive, $R\sim 1$, wherein a deviant nonmonotonic feature appears in the function $\omega(R)R^2$ (see Fig.~\ref{fig4}). The positive large-$R$ slope of this feature is related to the finite width of the Floquet-driven two-body resonance and the negative short-$R$ part is related to the finite value of $a_0$. This feature plays a central role in shifting both the three-body and the inelasticity parameters in our solution for the heavy particles which we now study. We also find that for all values of the drive parameters that we checked the imaginary part of $\omega(R)$ is always extremely small (at most of order $10^{-3}$). 
In fact, the results presented in the following subsection are obtained neglecting the imaginary part of $\omega(R)$. However, we have checked that including this imaginary piece does not lead to any noticeable difference.

\subsection{The heavy-atom problem}\label{HeavyAtom}

For sufficiently large mass ratios the motion of the heavy atoms can be assumed slow compared to both the light-atom motion and the drive frequency. The three-body problem thus reduces to the problem of two particles interacting by the isotropic potential $\omega(R)$. The radial Schr\"odinger equation for their relative motion in the angular-momentum channel $l$ reads
\begin{equation}\label{chi1}
  \left[-\frac{\partial^2}{\partial R^2}+\frac{l(l+1)}{R^2}+\frac{M}{2m}\omega(R)\right]\chi_l(R)=\frac{M}{2m}E\chi_l(R),
 \end{equation}
where $M\gg m$ is the heavy-atom mass, $E$ is the energy in units of $\hbar\Omega$, and we do not distinguish between the light-atom mass $m$ and the heavy-light reduced mass $\mu$. The relative discrepancy in doing this is of the order of $m/M\ll 1$. In fact, unless $l\gg 1$, keeping the centrifugal term compared to the interaction one in Eq.~(\ref{chi1}) exceeds this accuracy. However, even when $l=1$ the centrifugal barrier is important and leads to quantitative results. Indeed, in the undriven case for infinite light-heavy scattering length we have $\omega(R)=-C^2/R^2$ and Eq.~(\ref{chi1}) reduces to 
\begin{equation}\label{Undrivenchi1}
-\chi''_l+\frac{l(l+1)-C^2M/2m}{R^2}\chi_l=\frac{M}{2m}E\chi_l.
\end{equation}
The effective $1/R^2$ potential in this problem leads to the fall of a particle to the center for $l(l+1)-C^2M/2m<-1/4$ \cite{LandauLifshitz1987} and one can notice the interplay between the attraction due to the exchange of the light atom and centrifugal repulsion which can be dictated by the quantum statistics of the heavy atoms. In particular, the corresponding critical mass ratio in the channel with $l=1$, relevant to heavy identical fermions, equals $M/m=9/2C^2=13.990$. This number is to be compared to the exact (non-Born-Oppenheimer) threshold for the Efimov effect in this system $M/m=13.607$ \cite{Efimov,PetrovPRA2003}. One sees that the centrifugal term qualitatively and sufficiently quantitatively describes the onset of the Efimov effect and the role played by the quantum statistics of the heavy atoms. We will thus keep it also in our Born-Oppenheimer description of the driven system.

We remind that Eq.~(\ref{Undrivenchi1}) is a valid approximation to Eq.~(\ref{chi1}) for small and large $R$ also in the driven case. We thus describe its properties in more detail. The general solution of Eq.~(\ref{Undrivenchi1}) is expressed in terms of the Bessel functions as $\chi_l(R)\propto \sqrt{R}J_{\pm is_0}(\sqrt{ME/2m}R)$, where $s_0=\sqrt{C^2M/2m-(l+1/2)^2}$. The Efimovian regime (or the fall to the center) corresponds to real $s_0$. In this case these functions become oscillatory in the limit of small $R$ and the short-$R$ asymptote of the wave function can be written as 
\begin{equation}\label{chismallR}
\chi_l(R)\propto \sqrt{R}[e^{-\eta}(R/R_0)^{is_0}-e^{\eta}(R/R_0)^{-is_0}],
\end{equation}
where $R_0$ is the three-body parameter needed to fix the phase of the oscillations and $\eta>0$ is the inelasticity parameter, which reflects the fact that the amplitude of the incoming wave is larger than the amplitude of the outgoing one \cite{BraatenInelasticity}. 

We now go back to Eq.~(\ref{chi1}) and investigate the role played by the deviation of $\omega(R)$ from the pure $1/R^2$ scaling. First, let us assume that $E\ll 1$. This parameter is thus not important at distances $R\sim 1$ and we neglect it in Eq.~(\ref{chi1}). Then, by introducing the new coordinate $z=\ln R$ and transforming the wave function such that $\chi_l(R)=\exp(z/2)\Phi_l(z)$ Eq.~(\ref{chi1}) takes the form
\begin{equation}\label{Schrod1D}
-\frac{1}{M/2m}\Phi''_l(z)+W(z)\Phi_l(z)=\left[C^2-\frac{(l+1/2)^2}{M/2m}\right]\Phi_l(z)
\end{equation}
where $W(z)=e^{2z}\omega(e^z)+C^2$ is, up to the shift by $C^2$, what we plot in Fig.~\ref{fig4}. According to Eqs.~(\ref{omegalargeR}) and (\ref{omegasmallR}) the potential $W(z)$ decays exponentially for $|z|\rightarrow \infty$. We thus arrive at the simple one-dimensional barrier-reflection problem where the amplitudes of the left- and right-moving waves on the left of the barrier are related to each other by Eq.~(\ref{chismallR}) and calculating the corresponding amplitudes on the right of the barrier will give us the three-body and inelasticity parameters for the driven problem. Namely, in the new notations Eq.~(\ref{chismallR}) reads
\begin{equation}\label{Phileft}
\Phi_l(z)\propto e^{-\eta+is_0z+i\phi}-e^{\eta-is_0z-i\phi},
\end{equation}
where we have introduced the three-body phase $\phi=s_0\ln R_0$. For large $z$ (on the right of the barrier) $\Phi_l$ has the same form (\ref{Phileft}) but with three-body parameters $\phi_d$ and $\eta_d$ corresponding to the driven problem. The pair $\{\phi_d,\eta_d\}$ can be related to $\{\phi,\eta\}$ with the help of the reflection and transmission amplitudes defined by the scattering solution of Eq.~(\ref{Schrod1D}) characterized by the incident wave from the right, 
\begin{equation}\label{PhiScat}
\Psi_l(z)=\left\{
 \begin{array}{ll}
  {\cal{T}}e^{-i s_0 z} & z\ll 0 \\
  e^{-i s_0 z} + {\cal{R}} e^{i s_0 z} & z\gg 0.
 \end{array}\right.
 \end{equation}
The solution which satisfies the boundary condition in Eq.~\eqref{Phileft} is then the linear superposition of $\Psi_l$ and its complex conjugate,
\begin{equation}\label{Phil}
 \Phi_l(z)=\frac{1}{\cal{T}^*}\Psi_l^*(z)e^{-\eta+i\phi}-\frac{1}{\cal{T}}\Psi_l(z)e^{\eta-i\phi}.
\end{equation}
Substituting the large-$z$ asymptote of $\Psi_l$ given by Eq.~(\ref{PhiScat}) into Eq.~(\ref{Phil}) and identifying the inward and outward propagating waves we get
\begin{equation}\label{DrivenParameters}
    e^{\eta_d-i\phi_d}=e^{\eta-i\phi-i{\rm arg}{\cal{T}}}\sqrt{\frac{1-{\cal{R}}^*e^{-2(\eta-i\phi-i{\rm arg}{\cal{T}})}}
 {1-{\cal{R}}e^{2(\eta-i\phi-i{\rm arg}{\cal{T}})}}},
\end{equation}
which is the main result that exactly relates driven parameters $\{\phi_d,\eta_d\}$ to their undriven counterparts. 

\begin{figure}
    \centering
    \includegraphics[width=8.5cm]{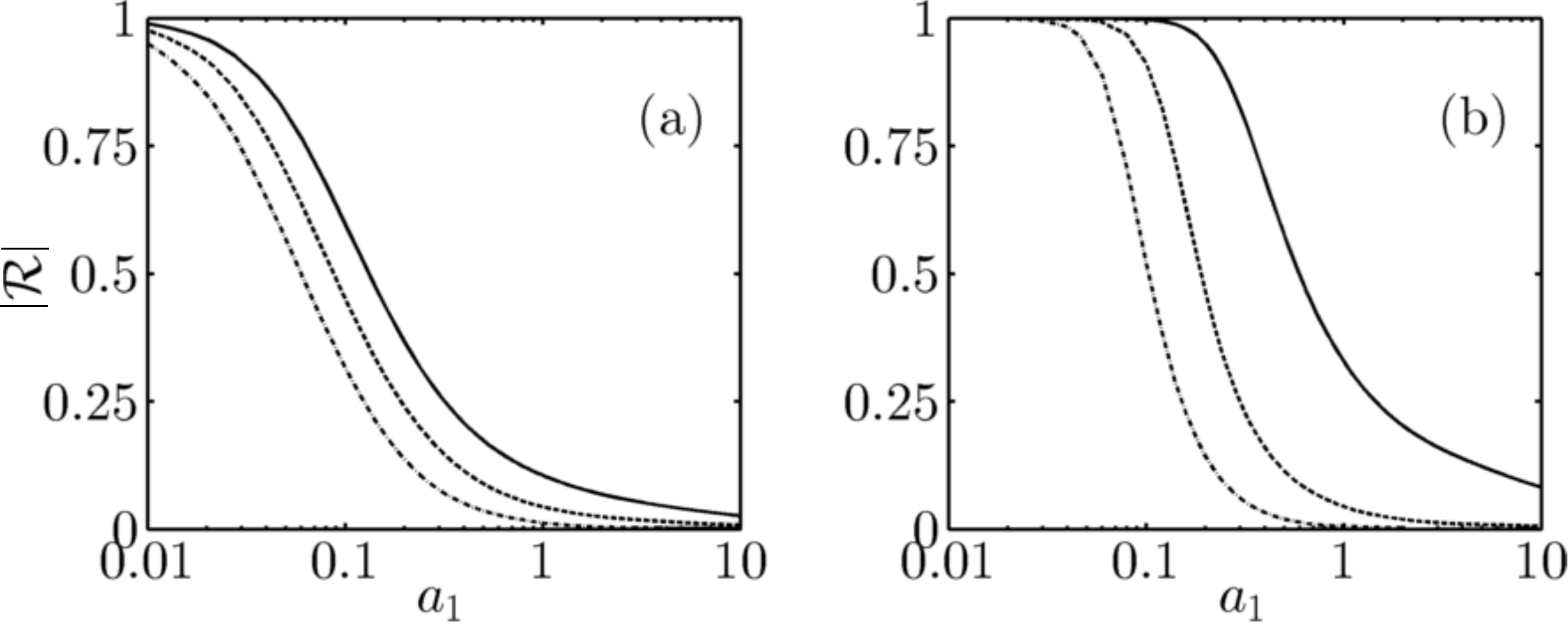}
    \caption{The reflection amplitude as a function of the drive parameter $a_1$ on the first resonance curve. In (a) we show the case of identical bosons where $l=0$. In (b) we show the case of identical fermions where $l=1$. The mass ratios have been chosen in (a) [(b)] to be 10, 20, and 40 [20, 40, and 80] distinguished by the solid, dashed, and dash-dotted curves, respectively.}
    \label{fig:R}
\end{figure}

The amplitudes $\cal{T}$ and $\cal{R}$ depend on the mass ratio, angular momentum, and, through the potential $W(z)$, on the position of the drive-parameters along the resonance curve. Accordingly, one can discuss a few limiting scenarios. We note that the reflection amplitude $\cal{R}$ increases with decreasing $a_1$ since $W(z)$ becomes stronger [see Fig.~(\ref{fig4})]. It also increases with decreasing $M/m$ since it reduces the effective mass and energy in Eq.~(\ref{Schrod1D}) thus making motion more quantum mechanical. The effective energy also decreases with $l$ leading to a significant increase of $|{\cal{R}}|$. In Fig.~\ref{fig:R} we illustrate this behavior by showing $|{\cal{R}}|$ as a function of $a_1$ ($a_0$ is changed accordingly to stay on the first Floquet resonance) for various values of $M/m$ in the cases $l=0$ (left panel) and $l=1$ (right panel).

\begin{figure}
    \centering
    \includegraphics[width=8.5cm]{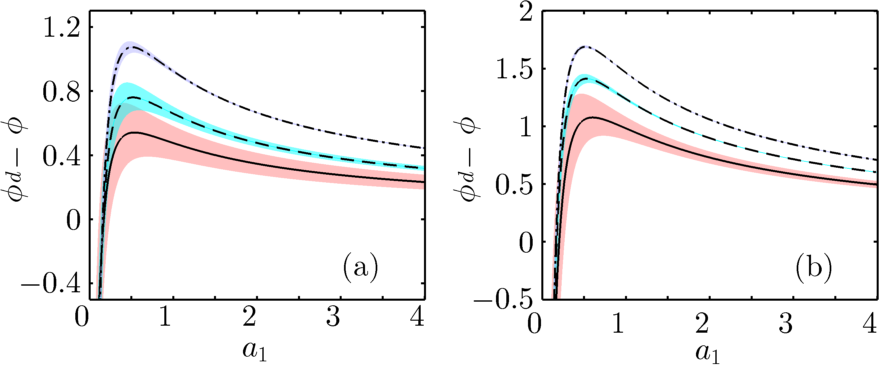}
    \caption{Shifting the three-body parameter as a function of the drive parameter $a_1$ (while staying on the first resonance position curve). In (a) we show the case of identical bosons with $l=0$. The solid, dashed, dashed-dotted, and dotted lines correspond, respectively, to mass ratios of 10, 25, 50, and 100. In (b) we show the case of identical fermions with $l=1$. The solid, dashed, and dashed-dotted lines correspond to mass ratios of 30, 60 and 90, respectively. The shaded region surrounding each line corresponds to the upper and lower bounds given by Eq.~\eqref{ThreeBodyShiftsUpperAndLower} whereas the lines themselves correspond to $\phi_d-\phi = {\rm arg}(T)$. }
    \label{fig:ThreeBodyParameter}
\end{figure}

Let us now discuss in more detail the case of small ${\cal{R}}$. Neglecting this parameter in Eq.~(\ref{DrivenParameters}) gives $\eta_d=\eta$ and $\phi_d=\phi+{\rm arg}\cal{T}$. This shift in the three-body phase can be explained classically as the motion of the fictitious particle governed by Eq.~(\ref{Schrod1D}) becomes faster (slower) in the region of negative (positive) $W(z)$. The particle passes the obstacle with a delay time proportional to $-{\rm arg}\cal{T}$. In Fig.~\ref{fig:ThreeBodyParameter} we plot ${\rm arg}\cal{T}$ as a function of $a_1$ for the same parameters as in Fig.~\ref{fig:R}. The positive shift for large $a_1$ is consistent with the dominance of the attractive part of $W(z)$ (see Fig.~\ref{fig4}).

For finite $\cal{R}$, the difference $\phi_d-\phi$ becomes $\phi$ and $\eta$ dependent and it can no longer be plotted as a single line in Fig.~\ref{fig:ThreeBodyParameter}. However, for small $|{\cal{R}}|$ this dependence is weak as can be seen directly by expanding Eq.~(\ref{DrivenParameters}) at small $|R|$. In fact, for $\eta=0$ one can derive exact inequalities
\begin{equation}
 {\rm arg}{\cal{T}}-{\rm arcsin}|{\cal{R}}|\leq \phi_d-\phi\leq {\rm arg}{\cal{T}}+{\rm arcsin}|{\cal{R}}|,\label{ThreeBodyShiftsUpperAndLower}
\end{equation}
valid for arbitrary $|{\cal{R}}|$. In Fig.~\ref{fig:ThreeBodyParameter} we show the bands (\ref{ThreeBodyShiftsUpperAndLower}) as shaded regions. The width of these regions increases with increasing $|{\cal{R}}|$ (compare with Fig.~\ref{fig:R}).


For small $a_1$ it is the repulsive part of $W(z)$ that becomes dominant. In addition to introducing a negative shift of the three-body phase it strongly increases the reflection amplitude. The obstacle becomes impenetrable for $|{\cal{R}}|\rightarrow 1$. In this limit $\eta_d\rightarrow 0$ and
\begin{equation}
 \phi_d=\frac{{\rm arg}{\cal{R}}}{2}+\frac{\pi}{2},
\end{equation}  
independent of the original $\phi$ and $\eta$. In order to estimate ${\rm arg}{\cal{R}}$ in the limit of small $a_1$ we recall that the repulsive tail of $W(z)$ is characterized by the length scale $R\sim R^*\propto 1/a_1^2$ [see Eq.~(\ref{omegalargeR})] which translates into $z\sim 2\ln(1/a_1)$ and, eventually, into the logarithmic scaling ${\rm arg}{\cal{R}}\approx 4s_0\ln a_1$ following from Eq.~(\ref{PhiScat}). This finding is consistent with the fact that near a narrow two-body resonance the effective range $R^*$ can play the role of the three-body parameter \cite{Petrov2004,GogolinMoraEgger2008,WangDIncaoEsry2011,SchmidtRathZwerger2012}. 


\begin{figure}
    \centering
    \includegraphics[width=8.5cm]{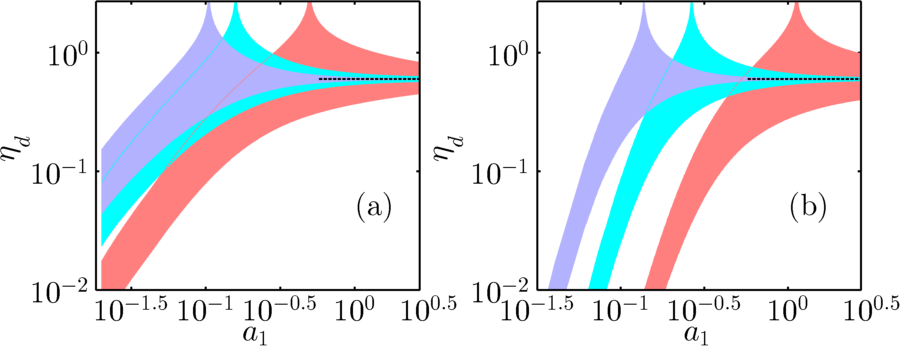}
    \caption{The inelasticity parameter as a function of the drive parameter $a_1$ (while staying on the first resonance curve) in the case of identical heavy bosons with $l=0$ (a) and identical heavy fermions with $l=1$ (b). The upper and lower boundaries of the shaded regions show $\eta_d^+$ and $\eta_d^-$, respectively, given by Eq.~\eqref{etapm}. In (a) [(b)] the mass ratios are 5, 20, and 40 [20, 40, and 80] for the red (rightmost), cyan (middle), and magenta (leftmost) regions. The undriven inelasticity parameter was set to 0.6 (shown as the black dashed line).}
    \label{fig:eta}
\end{figure}

We conclude this section by discussing the influence of the drive on the inelasticity parameter $\eta_d$. From Eq.~(\ref{DrivenParameters}) one can see that depending on $\phi$ this quantity can be anywhere inside the interval $\eta_d^{-}\leq\eta_d\leq \eta_d^{+}$, where
\begin{equation}\label{etapm}
 \eta_d^{\pm}=\frac{1}{2}\ln\frac{e^{2\eta}\mp|{\cal{R}}|}{|1\mp|{\cal{R}}|e^{2\eta}|}.
\end{equation}
In Fig.~\ref{fig:eta} we show the band $\{\eta_d^{-},\eta_d^{+}\}$ as a function of $a_1$ (staying on the first Floquet resonance) for $\eta=0.6$ and for $M/m=5$, $20$, and $40$ in the case $l=0$ (left panel) and $M/m=20$, $40$, and $80$ in the case $l=1$ (right panel). For large $a_1$ the reflection amplitude is small (see Fig.~\ref{fig:R}) and we get $\eta_d^{+}\approx\eta_d^{-}\approx \eta$. For small $a_1$ we have $|{\cal{R}}|\rightarrow 1$, $\eta_d^{-}\approx (1/2)(1-|{\cal{R}}|)\tanh\eta$, and $\eta_d^{+}\approx (1/2)(1-|{\cal{R}}|)\coth\eta$. Note that one can quite substantially reduce the inelasticity parameter by reducing $a_1$, particularly in the cases where $l=1$ because of the faster tendency $|{\cal{R}}|\rightarrow 1$ (see Fig.~\ref{fig:R}). Finally, in Fig.~\ref{fig:eta} one can also see peculiar divergences of $\eta_d^{+}$. They correspond to the vanishing denominator in Eq.~(\ref{etapm}) and indicate the possibility to reach the regime of total absorption $\eta_d=\infty$ by fine tuning $|{\cal{R}}|=e^{-2\eta}$ and $\phi={\rm arg}{\cal{T}}-{\rm arg}{\cal{R}}/2$ such that the denominator of Eq.~(\ref{DrivenParameters}) also vanishes. We note that whereas $\eta$ is fixed for a given Feshbach resonance, $\phi$ is tunable by changing the drive length $L_\Omega$ since in physical units $\phi=s_0\ln (R_0/L_\Omega)$.

\section{Conclusions and Discussion}\label{ConclusionsSection}

We have studied the two-body scattering problem in the zero-range approximation with a driven scattering length of the form $a(t)=a_0+2a_1\cos(\Omega t)$. We find a family of curves in the drive parameter space $\{a_0,a_1\}$ where the effective scattering length $a_{\rm eff}$ diverges. We also quantify the widths of these Floquet resonances in terms of the effective range parameter $R^*$. In the limit of weak drive (small $a_1$) these resonances occur when the molecular binding energy $\hbar^2/2\mu a_0^2$ equals an integer number of the drive quanta $\hbar\Omega$. In this limit the resonances are narrow which is manifested by diverging $R^*$. However, by increasing $a_1$ and moving along a Floquet-resonance curve one decreases $R^*$ thus changing the nature of the Floquet resonance from narrow to broad. Therefore, the Floquet driving gives us a tool for tuning the width of the two-body resonance. We expect that the resonant enhancement of the total cross section associated to these Floquet resonances could be measured via rethermalisation time scales in a cross dimensional relaxation experiment~\cite{MonroeCrossDimRelax1993,GoldwinCrossDimRelax2005,TangDipolarCrossDimRelax2015}.

In contrast to what one would expect, the Floquet resonances are not associated with enhanced absorption of drive quanta. In particular, exactly on resonance the two-body inelastic rate constant associated with this heating mechanism is proportional to the first power of the collision energy. Therefore, the two-body inelastic processes in a sufficiently cold Floquet-driven unitary gas become much weaker than the three-body ones. 

In order to analyse the influence of the drive on the three-body properties of an Efimovian system we have considered the problem of one light atom and two heavy identical bosons or fermions with the heavy-light interaction tuned to the Floquet resonance. In contrast to the well-known Efimov $1/R^2$ scaling for the heavy-heavy interaction induced by the exchange of the light atom, we find that in the driven case this potential develops a peculiar deviant feature when the heavy-heavy separation $R$ is of the order of the drive length $L_\Omega$. The long-range part of this feature can be associated to the finite width of the Floquet resonance parameterized by $R^*$ and its short-range part can be explained by the fact that the light-atom wave function adiabatically follows the trajectory $a(t)$. The main consequence of the appearance of this feature for the Efimov physics is that it partially reflects the incoming Efimov waves and introduces a phase shift for the transmitted part. We find that by moving along the Floquet-resonance curve one can control the effective inelasticity parameter and shift the three-body parameter. Although our approach uses the Born-Oppenheimer adiabatic assumption to treat the heavy-heavy-light problem, we see that the ability to control the three-body and inelasticity parameters increases with decreasing the mass ratio. This should stimulate further studies of the driven three-body problem with weaker mass imbalance without relying on the Born-Oppenheimer approximation.

Our theory is directly applicable to the Cs-Li ($M/m=22.1$) \cite{Pires2014,Tung2014,UlmanisEfimov2016} or Rb-Li ($M/m=14.45$) mixtures \cite{Zimmermann}. In particular, for the Cs-Cs-Li case the drive frequency $\Omega=2\pi \times 10$ kHz corresponds to the drive length $L_\Omega = 5600 a_{\rm Bohr}$ and the scattering length $a_0\approx L_\Omega$ is obtained for the detuning $\Delta B=-300$mG from the 842.829G Cs-Li Feshbach resonance characterized by the magnetic width of -58.21G and background scattering length -29.4$a_{\rm Bohr}$ \cite{CsLiFeshbach}. Under these conditions one can reach essentially any point on the first Floquet resonance curve by modulating the magnetic field with the amplitude below 300 mG. In particular, from Fig.~\ref{fig:ThreeBodyParameter} we see that modifying $a_1$ from 0.1 to 0.5$L_\Omega$ can change the three-body phase by more than $\pi/2$ which is enough to explore the essentials of the Efimov physics. We estimate that driving with $a_1\approx 0.1 L_\Omega$ can reduce the natural inelasticity parameter $\eta\approx 0.6$ \cite{UlmanisEfimov2016} for this mixture by a factor of three. 

These arguments suggest that positions of the three-body loss features and, more generally, the shapes of the loss curves can be significantly modified for a gas or mixture with Floquet-driven interactions. This topic requires a separate analysis beyond the scope of this paper. It may also provide an interesting path for future investigation, particularly in the context of the van der Waals universality of the three-body parameter in undriven cold atom systems~\cite{BerningerJulienneHutsonPRL2011,RoyInguscioModugnoPRL2013,BloomJinPRL2013,GreenePRL2012,WangJulienneNatPhys2014,HuangHutsonPetrovPRA2014}. 

Another direction for further research would be to consider an alternative drive. We have demonstrated the ability to control the width of the Floquet two-body resonance and three-body observables by using a simple sinusoidal drive as a tool. We believe that one may reach a more precise level of control over these observables or may even see qualitatively new physics in the case of a more complicated drive. Another frequency or other frequencies add other characteristic length scales to the problem, which should manifest itself in the modified shape of the deviant feature. These other frequencies may or may not be commensurate with the first one. In the former case one can still use the general formalism presented in Subsec.~\ref{TwoBodyFormalism}) dealing with the one-dimensional Floquet lattice. By contrast, the latter case can be represented in terms of a multi-dimensional Floquet lattice, which can exhibit nontrivial topological properties \cite{MartinRefaelHalperin2016}.

\section{Acknowledgements} We wish to thank J. Bohn and J. D'Incao for useful conversations. The research leading to these results received funding from the European Research Council (FP7/2007--2013 Grant Agreement No. 341197) and the European Union's Horizon 2020 research and innovation program under Grant Agreement No. 658311. H.L. acknowledges support by a Marie Curie Intra European Fellowship within the 7th European Community Framework Programme. We also acknowledge support by the IFRAF Institute.

\end{document}